\documentclass[conference]{IEEEtran}
\IEEEoverridecommandlockouts

\usepackage{multicol}
\usepackage{multirow}
\usepackage{float}
\usepackage{url}
\usepackage[table]{xcolor}
\definecolor{lightgray}{gray}{0.9}
\usepackage{booktabs}
\usepackage{tabularx}
\usepackage{makecell}
\usepackage{cite}
\usepackage{amsmath,amssymb,amsfonts}
\usepackage{algorithmic}
\usepackage{graphicx}
\usepackage{textcomp}
\usepackage{xcolor}
\usepackage{hyperref}
\def\BibTeX{{\rm B\kern-.05em{\sc i\kern-.025em b}\kern-.08em
    T\kern-.1667em\lower.7ex\hbox{E}\kern-.125emX}}
\begin{document}


\title{Cross-Lingual Parkinson's Disease Severity Assessment Using Pre-trained Speech Embeddings: A Multi-Class Evaluation

\thanks{This publication is part of the project Responsible AI for Voice Diagnostics (RAIVD) with file number NGF.1607.22.013 of the research programme NGF AiNed Fellowship Grants which is financed by the Dutch Research Council (NWO).}
}

\author{\IEEEauthorblockN{Simon Pals}
\IEEEauthorblockA{\textit{Centre for Language Studies} \\
\textit{Radboud University}\\
Nijmegen, the Netherlands \\
simon.pals@ru.nl \\ 0009-0008-0947-4305}
\and
\IEEEauthorblockN{Cristian Tejedor-Garcia}
\IEEEauthorblockA{\textit{Centre for Language Studies} \\
\textit{Radboud University}\\
Nijmegen, the Netherlands \\
cristian.tejedorgarcia@ru.nl \\ 0000-0001-5395-0438}
}

\maketitle

\begin{abstract}
Parkinson's disease (PD) often manifests through speech impairments, facilitating accessible, non-invasive, and cost-effective severity assessment for early diagnosis and progression tracking. Despite advances in speech foundation models (SFMs), their cross-lingual generalization for PD severity multi-class classification remains underexplored due to limited labeled data, a lack of explainable methods and variability across languages and datasets. In this work, we evaluate pre-trained embeddings from four state-of-the-art open-source SFMs across three datasets in zero-shot and k-shot cross-lingual settings for multi-class PD severity assessment. Our results show that pre-trained speech embeddings enable meaningful cross-lingual transfer, although performance is sensitive to dataset properties, preprocessing, and adaptation strategy. Misclassifications under these conditions related to inter-speaker variability and atypical speech patterns highlight the need for more robust feature extraction and modeling for PD severity assessment while emphasizing the importance of explainability for reliable clinical insights.\end{abstract}


\section{Introduction}

Parkinson's disease (PD), the second most prevalent neurodegenerative disorder worldwide, is characterized by progressive dopaminergic degeneration and is typically diagnosed after substantial irreversible damage \cite{Pagan2012-cq}. With a projected prevalence of more than 25 million cases by 2050 \cite{Bloem2021}, scalable and objective methods for assessing disease severity and progression are urgently needed \cite{su2025}.

Speech impairment affects up to 90\% of patients with PD \cite{Favaro2024} and manifests itself as hypophonia, monoprosody, reduced articulation rate, and imprecise phonation, often emerging early and evolving with disease progression and medication state \cite{morea2019speechimpairment}. Because speech production relies on fine motor control, its degradation reflects underlying neurophysiological dysfunction and provides a sensitive marker of disease severity \cite{MA2020speechimpairment}. Unlike traditional assessments based on neurological examination or imaging, which can be subjective, variable across raters, and resource-intensive \cite{beach2018,mayo-clinic-2024}, speech analysis offers a non-invasive, cost-effective, and scalable biomarker \cite{10767227}.

Beyond early detection, artificial intelligence (AI)-driven speech analysis provides objective and quantitative measures for tracking disease progression and treatment-related changes over time (e.g., early, mid, and late stages of PD), thereby complementing conventional clinical scales, which are inherently subjective \cite{chauhan2026ai}. Importantly, these tools are not intended to replace clinical expertise or guide treatment decisions; rather, they are designed to support the diagnostic process, particularly for less experienced clinicians, and to enable standardized, repeatable assessments \cite{10.3389/fnagi.2025.1638340}.

In parallel, developments in AI have strengthened the clinical relevance of speech-based biomarkers by enabling the detection of subtle deviations from healthy control (HC) speech patterns and supporting early identification of PD, including during prodromal or mild stages \cite{Orozco-Arroyave2020,VasquezCorrea2020}. This capability is particularly important for recruiting patients into clinical trials targeting early disease stages and for facilitating large-scale screening in settings where expert evaluation is not readily available. Besides, from a clinical perspective, cross-lingual modeling is essential for enabling scalable and accessible monitoring of disease progression across diverse linguistic populations \cite{Lim2025}.

AI has also driven the adoption of embedding-based approaches for speech-based PD assessment \cite{inventions10040048}, where fixed-dimensional acoustic embeddings are extracted from speech signals and used as input to downstream classifiers \cite{gelderen2024}. Compared to clinically motivated hand-crafted feature sets such as eGeMAPS \cite{egemaps}, embedding-based representations have demonstrated superior performance in both monolingual and cross-lingual scenarios \cite{FAVARO2023107559,laquatra2025bilingual,postma25_interspeech}. However, these gains often come at the cost of reduced explainability due to the opaque nature of deep neural architectures, motivating ongoing research into hybrid and explainable approaches \cite{Gomez2025interpretability,zhong25recapd}. Despite this progress, most existing studies remain confined to single-language settings, limiting their applicability in real-world clinical contexts characterized by substantial linguistic diversity.

Progress in speech-based PD severity classification, particularly in multi-class settings, remains constrained by the limited availability of publicly accessible datasets with reliable severity annotations. Most existing corpora are designed for binary PD versus HC classification, and only a small subset provides raw speech recordings labeled by disease severity \cite{gelderen2024}. Consequently, prior work often derives discrete severity classes by binning continuous clinical scores, such as the MDS-UPDRS-S \cite{KODALI2024101548,italianpvs2019} or the Hoehn and Yahr (HY) scale \cite{hoehn1967parkinsonism}, typically within monolingual frameworks.

From a clinical perspective, the ability to generalize across languages is essential, as many patients lack access to specialized centers or language-specific diagnostic tools. Consequently, cross-lingual PD severity assessment represents a critical step toward equitable, scalable, and clinically meaningful speech-based biomarkers that support both early detection and longitudinal monitoring across diverse populations \cite{hernandez2026adapting}.
Within this context, several studies have explored multi-class classification of PD speech patients with varying methodologies and performance. On the PC-GITA dataset, \cite{Kadiri_2023} employed SVM classifiers with acoustic features to discriminate HC, mild, and severe PD, reporting an accuracy of 63 ± 12\%, with a bias toward HC predictions. Using the same class configuration, \cite{KODALI2024101548} applied a multilayer perceptron (MLP) with clinically informed features and achieved 58\% accuracy. Other approaches include combining acoustic features with fine-tuned Whisper-derived character error rate features on the InhaPD dataset to classify HY stages 1–3, achieving 97.08\% accuracy in the OFF-medication condition \cite{Mondol2025}; training separate binary classifiers on PC-GITA to obtain an overall multi-class accuracy of 72\% \cite{OLIVEIRA2025109565}; and fine-tuning ImageNet-pretrained convolutional neural networks (CNNs) on log-mel spectrograms from the ItalianPVS dataset, yielding 91.8\% accuracy for HC, mild, and severe classes \cite{bioengineering11030295}.

In this work, we address two key methodological gaps in the literature. First, cross-lingual speech-based PD severity assessment using large pre-trained speech foundation models (SFMs) remains largely unexplored. These models learn language-agnostic acoustic representations from large multilingual corpora and are therefore well suited for PD assessment in scenarios where labeled clinical speech data are limited \cite{plantinga2025language}. Second, existing cross-lingual studies primarily focus on binary PD versus HC classification rather than multi-class severity stratification \cite{FAVARO2023107559,laquatra2025bilingual}. In contrast, monolingual studies using SFMs such as WavLM, Wav2Vec 2.0, Whisper, and OpenL3 have demonstrated strong performance for binary classification, with reported accuracies between 81\% and 87\% on datasets such as PC-GITA and NeuroVoz \cite{postma25_interspeech,La_Quatra_2024,purohit2025layerselection}. This gap limits the generalizability and external validity of current approaches. While binary classification is sufficient for screening, it does not support stage-specific assessment or longitudinal tracking of disease progression. 
%

To address these limitations, we present, to the best of our knowledge, the first systematic evaluation of multi-class PD severity assessment using pre-trained speech embeddings derived from open-source SFMs in a cross-lingual context, which constitutes the main novelty of this work.
Our contributions include 
(1) a systematic evaluation across three publicly available datasets and four SFMs,
(2) an open-source preprocessing pipeline for converting raw speech of any length into embeddings suitable for PD severity classification,
and (3) a detailed post-hoc analysis of misclassified and hard cases to explain model behavior and cross-lingual generalization constraints. We also highlight key methodological pitfalls and provide practical guidance for the research community.
In particular, our research questions (RQs) are: \textit{RQ1}. To what extent can multi-class PD-severity be classified with pre-trained speech embeddings from SFMs in a cross-lingual context, under zero-shot and k-shot conditions?, 
and \textit{RQ2}. Which patterns from post-hoc analysis of hard-to-classify cases reveal limitations in PD severity classification with pre-trained speech embeddings?


\section{Method}\label{sec:method}

\subsection{Datasets}
Three public speech datasets of different languages were selected for this study: ItalianPVS (Italian) \cite{italianpvs2019}, MDVR-KCL (English) \cite{mdvrkcl2019}, and PC-GITA (Colombian Spanish) \cite{pcgita2014}. They were chosen due to their widespread use in prior PD research, their shared reading-speech task (where participants read phonetically balanced sentences or passages), and the inclusion of PD speakers across a range of MDS-UPDRS-S speech severity scores, thereby enabling fair and meaningful cross-dataset comparisons.

Following \cite{KODALI2024101548, Kadiri_2023, bioengineering10080984}, we performed binning of the severity scores to formulate a three-class multi-class classification problem. Participants diagnosed with PD were categorized as \textit{Mild} for scores of 0 (no clinicallly observable speech impairment) or 1, and as \textit{Severe} for scores of 2 or higher, whereas independently identified healthy participants formed the \textit{HC} class. Accordingly, a score of 0 indicates normal speech in the presence of PD, and the HC class indicates the absence of PD.


\begin{figure*}[ht!]
    \centering
    \includegraphics[width=0.9\linewidth]{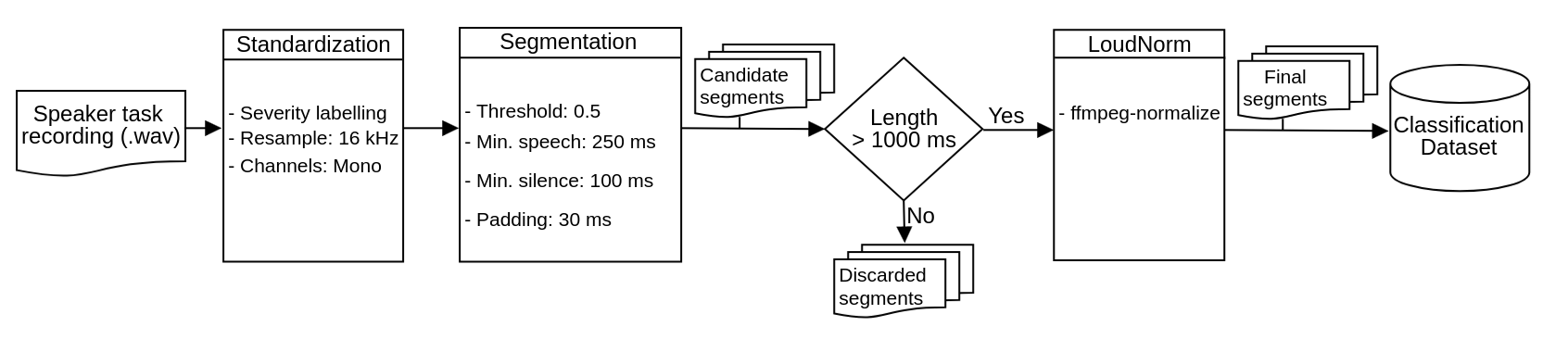}
    \vspace{-0.3cm}
    \caption[Preprocessing]{Flowchart of the preprocessing steps for each speech file of the source datasets.}
    \label{fig:preproc_prot}
\end{figure*}

After labelling the speakers with a severity score, the audio recordings in the datasets were preprocessed following the protocol visualized in Figure \ref{fig:preproc_prot}. 
First, label standardization was performed using severity labels derived from the MDS-UPDRS-S scores provided in the metadata of the datasets. Second, recordings were segmented to obtain comparable speech samples of similar duration. 
To address variability in recording lengths and ensure that the extracted embeddings represent approximately the same amounts of information, we segmented the speech files using the Silero VAD model \cite{SileroVAD} to extract comparable speech segments of similar duration, thereby balancing the input samples for subsequent processing and classification. Segments shorter than 1000 milliseconds were discarded from the dataset.
Third, loudness normalization was applied to reduce variability in recording conditions.
Each dataset was divided into K speaker-disjoint and Stratified folds, to train the models using speaker independent cross-validation and ensure speakers of all classes are represented in each fold by at least one speaker. In our work, we chose K=4 to handle the 'Severe' class imbalance.

\begin{figure}[ht!]
    \centering
    \includegraphics[width=1\linewidth]{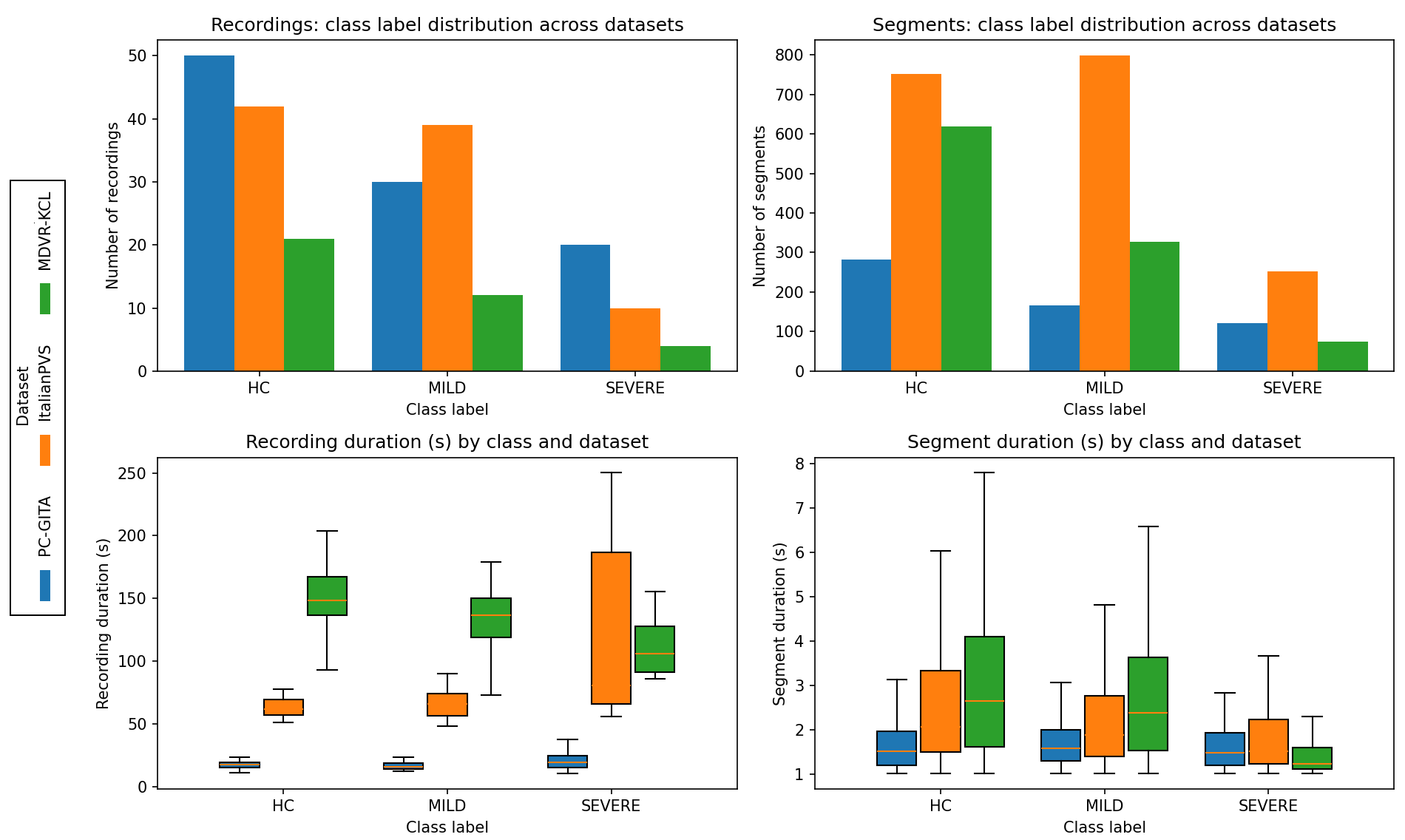}
    \caption[Dataset Properties]{Distribution of original and segmented speech files.}
    \label{fig:dataset_properties}
\end{figure}

Figure \ref{fig:dataset_properties} shows that segmentation (right) preserves the original distribution (left); however, the Severe PD class remains underrepresented, resulting in class imbalance.




\subsection{Modeling}

We employed four open-source, state-of-the-art SFMs to extract embeddings for PD classification, using language-specific fine-tuned versions retrieved from Hugging Face.\footnote{\scriptsize{\url{https://huggingface.co/models}, last visited: 2026-06-16.}} The models include Wav2Vec 2.0 XLS-R \cite{babu22_interspeech}, a multilingual self-supervised ASR model pretrained via contrastive learning; WavLM \cite{wavlm2021chen}, trained with denoising and masked prediction objectives; Whisper-v3 \cite{radford2022whisper}, a weakly supervised multilingual ASR and speech translation model; and OpenL3 \cite{cramer2019openl3}, an audio-visual self-supervised embedding model. The extracted embeddings were used to train four multi-class classifiers commonly adopted in prior work: ERT, SVM, XGB, and MLP. Features were standardized using scikit-learn’s \cite{scikit-learn} StandardScaler, and hyperparameters were optimized via GridSearchCV over a constrained search space. The MLP was trained with AdamW and early stopping (maximum 25 epochs, patience = 5).

\subsection{Experimental Procedure}

\begin{figure*}[ht!]
    \centering
    \includegraphics[width=0.9\linewidth]{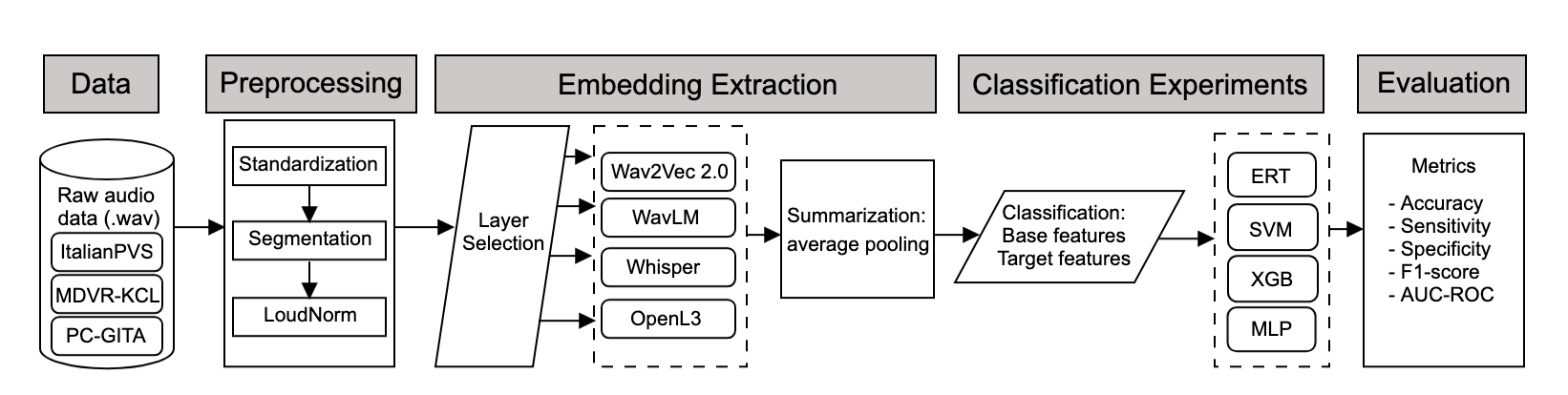}
    \vspace{-0.6cm}
    \caption[Experimental Pipeline]{Overview of the experimental steps.}
    \label{fig:pipeline}
\end{figure*}

In this study, a systematic evaluation was conducted under 16 classification pipelines, each defined by a combination of an SFM and a ML classifier. The code\footnote{\scriptsize{\url{https://github.com/simon-hjp/pd-speech-embedding-severity-classification},last visited: 2026-09-15.}} was run on a PC with a Ryzen 9950X CPU, an Nvidia 4060 Ti GPU with 16 GB VRAM, and 128 GB system RAM. The overall procedure is illustrated in Figure \ref{fig:pipeline}. 
From left to right, the raw speech recordings from the three datasets were preprocessed in the three main steps (standardization, segmentation and normalization) already described in Figure \ref{fig:dataset_properties}.

After preprocessing, embeddings were extracted from a specific encoder layer of each SFM, as layer choice can substantially influence performance on downstream tasks \cite{purohit2025layerselection}. For each dataset–SFM pair, the layer resulting in the highest validation accuracy for that dataset was selected. Frame-level embeddings were then aggregated via average pooling \cite{Klempir2025pooling} to obtain a fixed-length segment representation, whose dimensionality corresponds to the encoder’s embedding size. Table \ref{tab:encoder_layers} shows the encoder layers chosen for each dataset-SFM pair. OpenL3 is not included in this layer-selection analysis because it directly outputs audio embeddings rather than exposing a comparable set of encoder layers for selection.
These representations were used to train the ML classifiers. Following hyperparameter optimization, models were refitted on the combined training and validation data and evaluated on a held-out test set under speaker-disjoint K-fold cross-validation.

\begin{table}[ht!]
\centering
\caption{Selected encoder layers for each model and dataset}
\begin{tabularx}{\linewidth}{lXXX}
\toprule
\textbf{Dataset} 
& \textbf{Wav2vec 2.0} & \textbf{WavLM} & \textbf{Whisper} \\
\midrule
PC-GITA & 9 & 12 & 32 \\
ItalianPVS & 7 & 8 & 12 \\
MDVR-KCL & 3 & 6 & 7 \\
\bottomrule
\end{tabularx}
\label{tab:encoder_layers}
\end{table}

Finally, the performance of the assessment of PD severity was evaluated using accuracy and macro-averaged precision, recall, and F1-score, together with macro one-vs-rest ROC-AUC. Implementations in scikit-learn were used for each metric. First, performance was evaluated in a zero-shot setting, training on only the base language and testing on both base and target languages; and then in k-shot, where the classification head was retrained on a combined set of embeddings in both the source and target languages. This combined set was created by extending it with \textit{k} target-language segments per-class from the target-language training fold. The classification pipeline was then tested on the base language and the target language. The value of \textit{k} was set to the minimum per-class number of available segments observed across all folds and datasets (49), to ensure the same k-shot setting in all experiments.

\begin{table*}[ht!]
    \centering
    \scriptsize
    \caption[Cross-lingual results overview]{Classification performance for each source $\rightarrow$ target dataset pair with the highest F1-score of the target language}
    \vspace{-0.3cm}
    \setlength{\tabcolsep}{3pt}
    \begin{tabular}{|c|c|cccccccc|}
    \hline
    Target & Source & Pipeline & Condition & Language & Accuracy (\%) & Precision (\%) & Recall (\%) & F1 (\%) & ROC-AUC (\%) \\
    \hline
     & & WavLM + XGB & Zero-shot & Target & 58.14 $\pm$ 8.50 & 58.18 $\pm$ 7.54 & 57.73 $\pm$ 7.64 & 53.69 $\pm$ 6.38 & 75.03 $\pm$ 6.63 \\
    \rowcolor{lightgray}
    \cellcolor{white} & \cellcolor{white} & WavLM + SVM & Adapted (k-shot) & Target & 81.73 $\pm$ 6.36 & 75.02 $\pm$ 7.99 & 72.38 $\pm$ 2.52 & 71.02 $\pm$ 4.74 & 86.25 $\pm$ 8.74 \\
    & & WavLM + SVM & Zero-shot & Source & 79.19 $\pm$ 6.09 & 64.64 $\pm$ 15.81 & 60.62 $\pm$ 8.81 & 60.93 $\pm$ 11.95 & 88.30 $\pm$ 2.07 \\
     \rowcolor{lightgray}
    \cellcolor{white} & \multirow{-4}{*}{\cellcolor{white} \textbf{MDVR-KCL}} & WavLM + SVM & Adapted (k-shot) & Source & 78.07 $\pm$ 7.42 & 66.44 $\pm$ 13.57 & 60.05 $\pm$ 7.35 & 61.01 $\pm$ 10.09 & 89.00 $\pm$ 2.59 \\
     \cline{2-10}
    \cellcolor{white} & & WavLM + ERT & Zero-shot & Target & 43.85 $\pm$ 2.88 & 42.26 $\pm$ 4.66 & 43.86 $\pm$ 8.08 & 37.75 $\pm$ 4.65 & 63.35 $\pm$ 7.05 \\
     \rowcolor{lightgray}
    \cellcolor{white} & \cellcolor{white} & WavLM + ERT & Adapted (k-shot) & Target & 77.38 $\pm$ 7.10 & 70.76 $\pm$ 4.81 & 72.96 $\pm$ 3.99 & 67.66 $\pm$ 6.75 & 87.17 $\pm$ 4.05 \\
    \cellcolor{white} & & WavLM + ERT & Zero-shot & Source & 59.79 $\pm$ 2.60 & 57.96 $\pm$ 3.81 & 49.20 $\pm$ 2.05 & 48.96 $\pm$ 2.56 & 73.62 $\pm$ 7.27 \\
     \rowcolor{lightgray}
    \multirow{-8}{*}{\cellcolor{white} ItalianPVS} & \multirow{-4}{*}{\cellcolor{white} PC-GITA} & WavLM + ERT & Adapted (k-shot) & Source & 60.01 $\pm$ 5.67 & 57.42 $\pm$ 9.19 & 50.02 $\pm$ 5.29 & 50.02 $\pm$ 6.64 & 74.65 $\pm$ 5.81 \\
    \hline
    \cellcolor{white} & & Whisper + MLP & Zero-shot & Target & 59.91 $\pm$ 2.23 & 48.89 $\pm$ 8.71 & 50.71 $\pm$ 9.63 & 47.08 $\pm$ 7.90 & 73.73 $\pm$ 4.76 \\
     \rowcolor{lightgray}
    \cellcolor{white} & \cellcolor{white} & Whisper + SVM & Adapted (k-shot) & Target & 78.80 $\pm$ 5.64 & 65.80 $\pm$ 11.05 & 64.00 $\pm$ 7.61 & 63.98 $\pm$ 9.34 & 86.04 $\pm$ 4.15 \\
     & & Whisper + SVM & Zero-shot & Source & 82.09 $\pm$ 7.75 & 69.73 $\pm$ 11.05 & 67.75 $\pm$ 4.55 & 65.40 $\pm$ 6.28 & 83.35 $\pm$ 7.17 \\
     \rowcolor{lightgray}
    \cellcolor{white} & \multirow{-4}{*}{\cellcolor{white} ItalianPVS} & Whisper + SVM & Adapted (k-shot) & Source & 82.53 $\pm$ 7.75 & 70.26 $\pm$ 11.19 & 68.87 $\pm$ 5.05 & 66.18 $\pm$ 6.71 & 82.82 $\pm$ 7.98 \\
     \cline{2-10}
     & & OL3 + SVM & Zero-shot & Target & 58.34 $\pm$ 7.13 & 54.65 $\pm$ 6.60 & 63.07 $\pm$ 2.09 & 52.71 $\pm$ 5.09 & 76.29 $\pm$ 3.67 \\
     \rowcolor{lightgray}
    \cellcolor{white} & \cellcolor{white} & W2V + XGB & Adapted (k-shot) & Target & 74.06 $\pm$ 4.27 & 66.70 $\pm$ 3.39 & 71.81 $\pm$ 3.60 & 66.91 $\pm$ 3.36 & 86.88 $\pm$ 1.64 \\
     & & W2V + XGB & Zero-shot & Source & 62.00 $\pm$ 5.94 & 59.00 $\pm$ 6.63 & 55.30 $\pm$ 4.36 & 56.10 $\pm$ 5.04 & 76.28 $\pm$ 3.93 \\
    \rowcolor{lightgray}
    \multirow{-8}{*}{\cellcolor{white} MDVR-KCL} & \multirow{-4}{*}{\cellcolor{white} \textbf{PC-GITA}} & W2V + XGB & Adapted (k-shot) & Source & 62.56 $\pm$ 6.67 & 58.69 $\pm$ 6.98 & 56.56 $\pm$ 6.02 & 57.00 $\pm$ 6.23 & 76.27 $\pm$ 3.21 \\
    \hline
     &  & OL3 + MLP & Zero-shot & Target & 38.52 $\pm$ 4.65 & 44.89 $\pm$ 8.52 & 35.09 $\pm$ 3.29 & 35.21 $\pm$ 3.08 & 49.52 $\pm$ 8.02 \\
    \rowcolor{lightgray}
    \cellcolor{white} & \cellcolor{white} & W2V + SVM & Adapted (k-shot) & Target & 58.19 $\pm$ 2.61 & 58.02 $\pm$ 5.35 & 56.13 $\pm$ 3.35 & 56.46 $\pm$ 4.00 & 75.47 $\pm$ 5.45 \\
     &  & W2V + SVM & Zero-shot & Source & 81.86 $\pm$ 9.84 & 64.37 $\pm$ 7.93 & 69.37 $\pm$ 6.87 & 65.04 $\pm$ 6.52 & 82.79 $\pm$ 10.20 \\
    \rowcolor{lightgray}
    \cellcolor{white} & \multirow{-4}{*}{\cellcolor{white} \textbf{ItalianPVS}} & W2V + SVM & Adapted (k-shot) & Source & 81.30 $\pm$ 10.93 & 64.68 $\pm$ 9.40 & 69.74 $\pm$ 8.51 & 65.21 $\pm$ 7.41 & 85.60 $\pm$ 9.94 \\
     \cline{2-10}
     &  & OL3 + MLP & Zero-shot & Target & 49.66 $\pm$ 4.92 & 57.37 $\pm$ 7.47 & 44.40 $\pm$ 4.81 & 42.39 $\pm$ 4.66 & 64.79 $\pm$ 3.90 \\
    \rowcolor{lightgray} 
    \cellcolor{white} & \cellcolor{white} & WavLM + SVM & Adapted (k-shot) & Target & 57.01 $\pm$ 5.80 & 54.70 $\pm$ 4.99 & 52.91 $\pm$ 5.62 & 52.88 $\pm$ 4.75 & 72.80 $\pm$ 4.81 \\
     &  & WavLM + SVM & Zero-shot & Source & 79.19 $\pm$ 6.09 & 64.64 $\pm$ 15.81 & 60.62 $\pm$ 8.81 & 60.93 $\pm$ 11.95 & 88.30 $\pm$ 2.07 \\
    \rowcolor{lightgray}
    \multirow{-8}{*}{\cellcolor{white} PC-GITA} & \multirow{-4}{*}{\cellcolor{white} MDVR-KCL} & WavLM + SVM & Adapted (k-shot) & Source & 78.51 $\pm$ 6.43 & 63.41 $\pm$ 14.05 & 58.68 $\pm$ 6.49 & 59.25 $\pm$ 9.36 & 88.59 $\pm$ 1.09 \\
    \hline
    \end{tabular}
    \label{tab:crosslingual_results_summary}
\end{table*}

\section{Results and Discussion}\label{sec:results}

\subsection{Cross-lingual Analysis}\label{sec:crosslingual_results}

In order to answer RQ1, we first evaluated the impact of using zero-shot and k-shot cross-lingual adaptation on multi-class PD-severity assessment, for each dataset. Additionally, we evaluated the effect per severity class within each dataset.
Table \ref{tab:crosslingual_results_summary} reports the detailed performance of the best-performing configuration for each source-target pair, selected by target-language F1-score.
Table \ref{tab:f1_transfer_effects} summarizes the target-language F1-score changes and the corresponding source-language F1-changes for each source $\rightarrow$ target dataset pair. In this table, F1$^{Target}_{0-shot}$ and F1$^{Target}_{k-shot}$ denote the best zero-shot and k-shot target language F1-scores for the same source $\rightarrow$ target dataset pair, respectively. $\Delta$F1$^{Target}$ is computed as the difference between these variables, while $\Delta$F1$^{Source}$ denotes the source-language F1 change for the selected k-shot pipeline, computed with and without adding target-language samples during training. Positive values indicate an F1 increase after adaptation, negative values indicate the reverse. Target-language F1 increased in all cases, while source-language F1 remained comparatively stable.
We also observe that the performance varies with classifier, embedding type, and metric, but the dominant factor is the target dataset rather than the source. ItalianPVS shows the highest scores, followed by MDVR-KCL, while PC-GITA remains the most challenging. Therefore, we do not designate an overall best pipeline, since there is no configuration that consistently performs best across experimental conditions.

\begin{table}[ht!]
\centering
\scriptsize
\caption{Target- and source-language F1-score changes for each\\ source $\rightarrow$ target dataset pair}
\vspace{-0.2cm}
\setlength{\tabcolsep}{3.5pt}
\begin{tabular}{llrrrr}
\toprule
Target & Source & F1$^{Target}_{0-shot}$ & F1$^{Target}_{k-shot}$ & $\Delta$F1$^{Target}$ & $\Delta$F1$^{Source}$ \\
\midrule
ItalianPVS & MDVR-KCL & 53.69 & 71.02 & +17.33 & +0.08 \\
ItalianPVS & PC-GITA & 37.75 & 67.66 & +29.91 & +1.06 \\
MDVR-KCL & ItalianPVS & 47.08 & 63.98 & +16.90 & +0.78 \\
MDVR-KCL & PC-GITA & 52.71 & 66.91 & +14.20 & +0.90 \\
PC-GITA & ItalianPVS & 35.21 & 56.46 & +21.25 & +0.17 \\
PC-GITA & MDVR-KCL & 42.39 & 52.88 & +10.49 & -1.68 \\
\bottomrule
\end{tabular}
\label{tab:f1_transfer_effects}
\end{table}

\begin{figure}[ht!]
     \centering
     \includegraphics[width=1\linewidth]{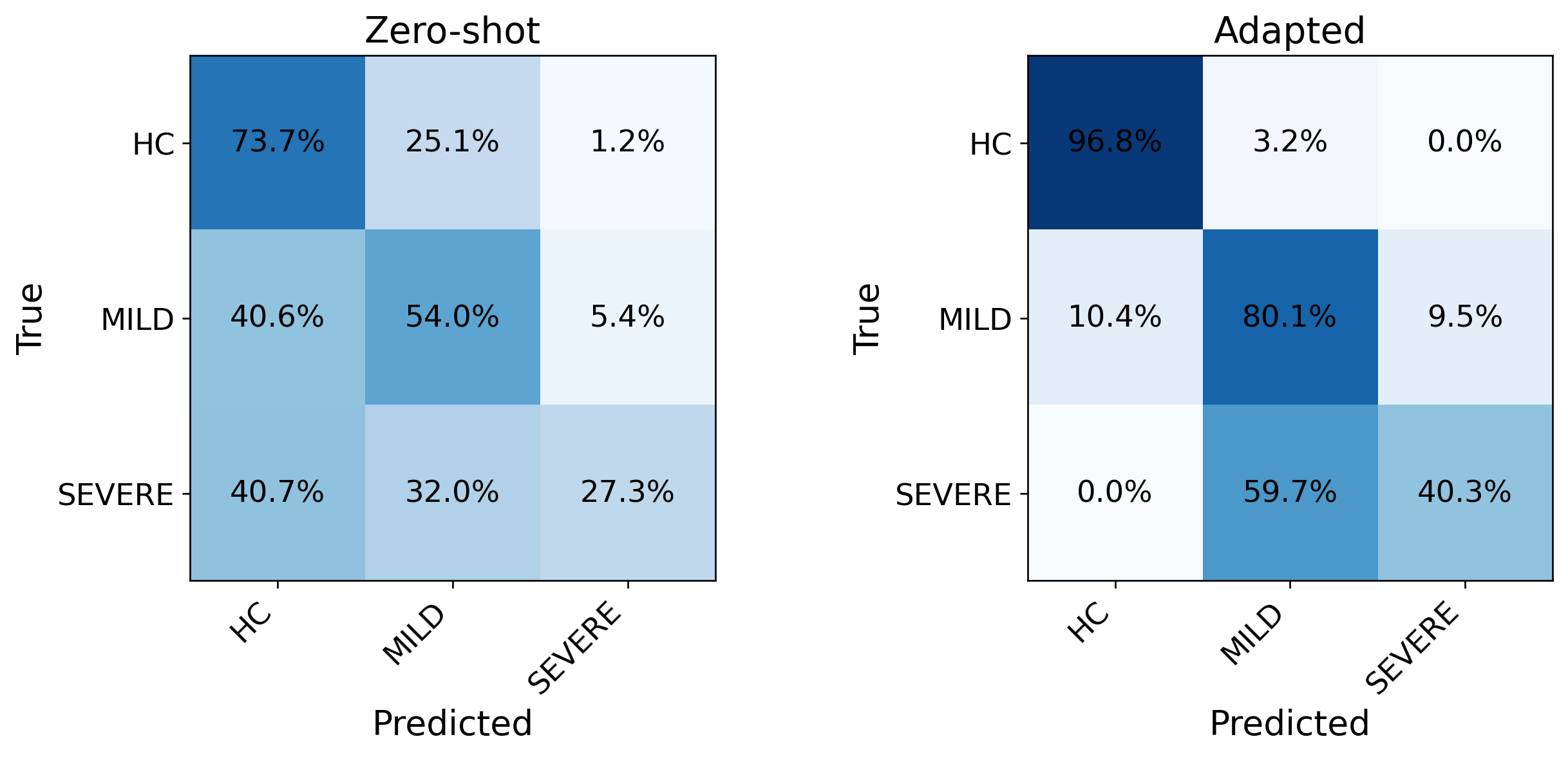}
     \vspace{-0.3cm}
     \caption[ItalianPVS, Cross-lingual Confusion Matrix]{Confusion matrix of the ItalianPVS dataset.}
     \label{fig:confmat_italianpvs_cross}
 \end{figure}

In particular, for target language ItalianPVS, using MDVR-KCL as source outperforms PC-GITA as source across all metrics except ROC-AUC, and k-shot adaptation consistently improves target performance with minimal impact on the source language. Zooming into the three severity classes, Figure \ref{fig:confmat_italianpvs_cross} shows that after k-shot adaptation, the Severe class is predominantly confused with the Mild class (59.7\% of cases), while the other two classes achieve high recognition rates (80.1\% and 96.8\%, respectively).

\begin{figure}[ht!]
     \centering
     \includegraphics[width=1\linewidth]{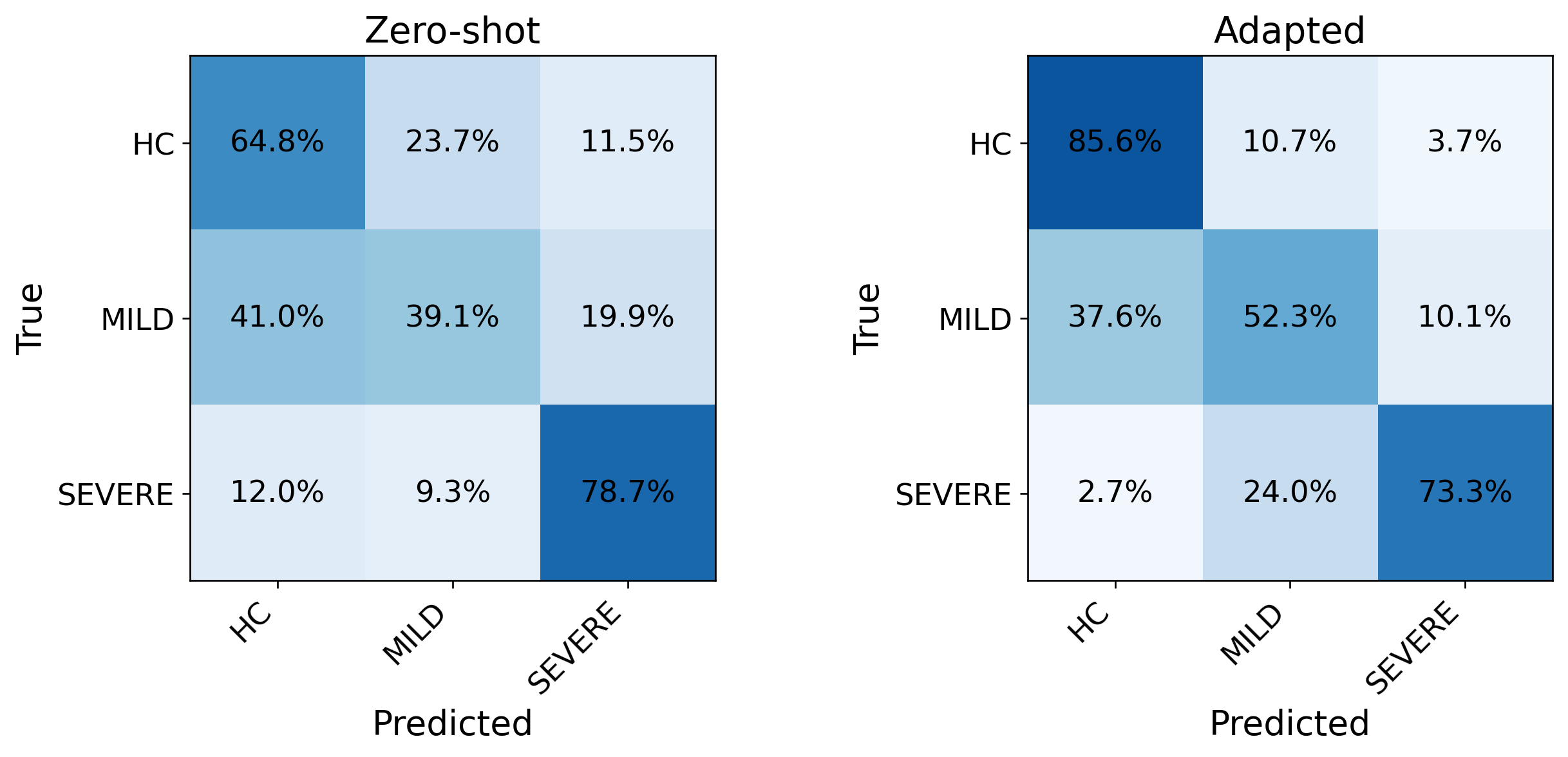}
     \vspace{-0.3cm}
     \caption[MDVR-KCL, Cross-lingual Confusion Matrix]{Confusion matrix of the MDVR-KCL dataset.}
     \label{fig:confmat_mdvrkcl_cross}
 \end{figure}

In the case of MDVR-KCL as the target, PC-GITA as the source yields the best performance (except for accuracy), and the k-shot adaptation improves all metrics over the zero-shot model while maintaining stable source-language performance. Figure \ref{fig:confmat_mdvrkcl_cross} reflects improved HC and Mild PD recognition after adaptation, at the cost of a slight reduction in Severe PD recall, with Severe cases more frequently misclassified as Mild.

\begin{figure}[ht!]
     \centering
     \includegraphics[width=1\linewidth]{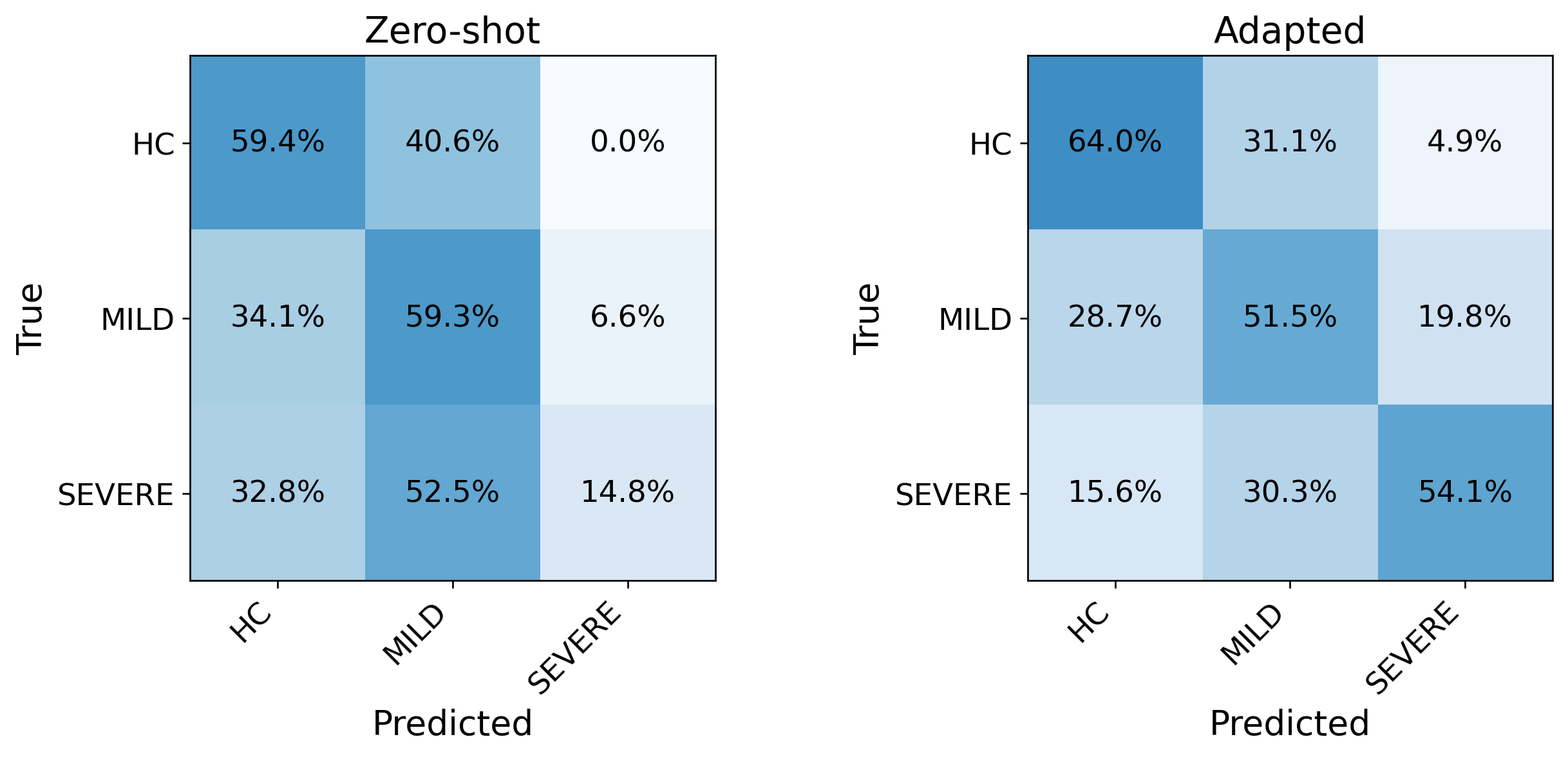}
     \vspace{-0.3cm}
     \caption[PC-GITA, Cross-lingual Confusion Matrix]{Confusion matrix for the PC-GITA dataset.}
     \label{fig:confmat_pcgita_cross}
 \end{figure}

Finally, for PC-GITA as target, ItalianPVS as source outperforms MDVR-KCL, and k-shot adaptation improves all metrics compared to zero-shot; however, PC-GITA remains the most challenging dataset, with only ROC-AUC exceeding 60\%. Figure \ref{fig:confmat_pcgita_cross} shows improved Severe class assessment in relation to the other two classes, although it remains moderately confused with the Mild class. A similar confusion pattern of HC being predicted as Mild is observed in both settings.

Answering RQ1, the performance gap between zero-shot and k-shot pipelines is markedly larger on the target dataset than on the source dataset. The analysis of confusion matrices indicates that zero-shot models predominantly misclassify Mild and Severe PD, and occasionally Severe PD as HC, whereas adapted models mainly confuse Mild with Severe/HC, consistent with observations by \cite{KODALI2024101548}. 
Increasing the representation of Severe PD during adaptation improved its assessment, albeit sometimes at the expense of other classes.
These findings suggest that SFM embeddings encode cross-lingual PD-severity information; however, effective exploitation of these representations requires exposure to target-language samples \cite{FAVARO2023107559}, which does not degrade source-language performance, as metrics and variability remain stable.

\subsection{Post-hoc Analysis}\label{sec:posthoc_results}

\begin{figure}[ht!]
    \centering
    \includegraphics[width=1\linewidth]{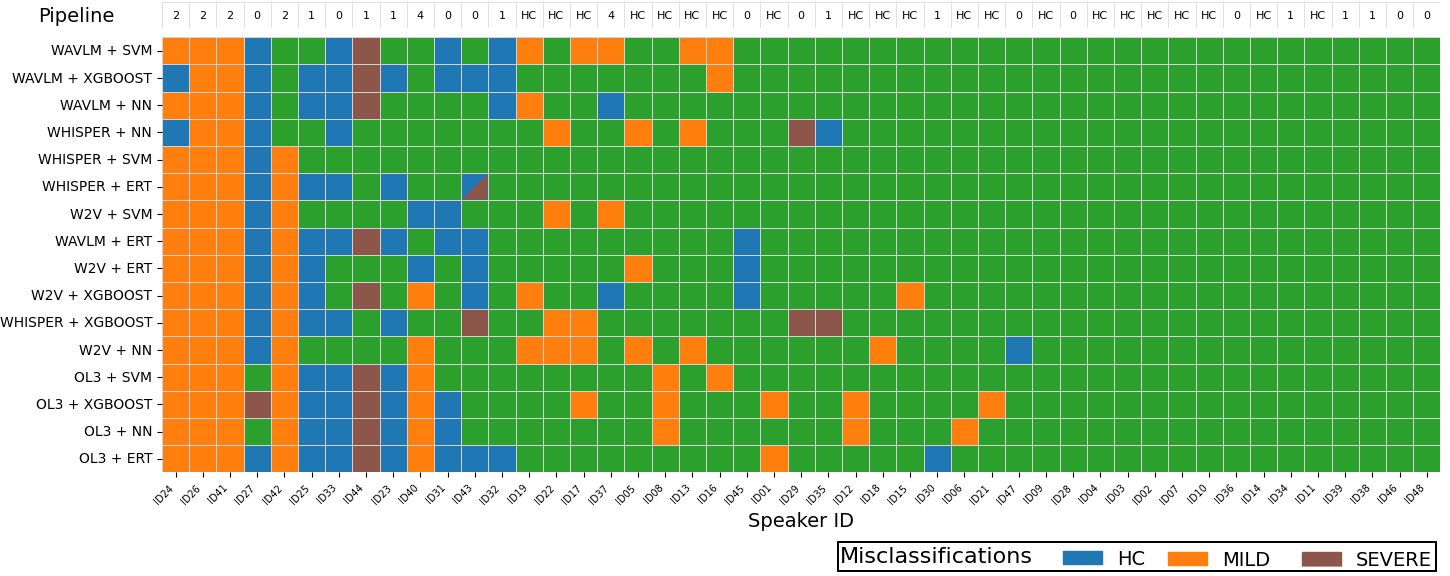}
    \vspace{-0.5cm}
    \caption[ItalianPVS, cross-lingual condition speaker hardness grid]{Speaker hardness grid of the ItalianPVS dataset.}
    \label{fig:posthoc_italianpvs_cross}
\end{figure}

To answer RQ2, we first analyzed inter-speaker error patterns quantitatively across all pipelines and datasets. Due to space constraints, only the results for the ItalianPVS dataset are shown in Figure \ref{fig:posthoc_italianpvs_cross}, where the grid displays speakers (columns) and pipelines (rows), ordered by speaker hardness and segment-level F1-score, respectively. A speaker is labeled as \textit{hard} if at least 50\% of their segments are misclassified by a given pipeline (and coloured according to most-frequent misclassification), and easy (green) otherwise, inspired by \cite{postma25_interspeech}. The top row indicates the ground-truth MDS-UPDRS-S speech scores, or HC where applicable. 


Across datasets, a subset of speakers with MDS-UPDRS-S score 2 (Severe PD) were consistently hard cases, with most segments misclassified by all pipelines, indicating speaker-specific factors that limit reliable severity assessment. This concentration near the Mild–Severe boundary suggests label ambiguity and motivates a potential intermediate category (e.g., Moderate PD), consistent with the clinical interpretation of score 2 \cite{italianpvs2019,OLIVEIRA2025109565}.
Besides this, there is variability between pipelines, likely due to limited target-language data that different pipelines handle unevenly. This highlights the importance of pipeline selection in cross-lingual PD-severity assessment, especially under low-resource conditions \cite{postma25_interspeech,purohit2025layerselection}.

Because Table \ref{tab:crosslingual_results_summary} ranks pipelines by segment-level F1, the best segment-level model does not necessarily yield the highest speaker-level accuracy, as shown in Figure \ref{fig:posthoc_italianpvs_cross}.
This discrepancy reflects the differing aggregation of errors at the speaker level. Therefore, for clinical decision support, pipeline selection should prioritize speaker-level performance or integrate both evaluation levels, as speaker-level aggregation better reflects patient-level decision-making in clinical PD assessment \cite{tougui2025crosslingualmultigranularityframeworkinterpretable}.



Finally, qualitatively listening to the recordings of hard speakers identified in the grids revealed speech-related insights about misclassification patterns. Although PD speech is typically associated with hoarseness, reduced speech intensity and rate, monopitch, and imprecise articulation \cite{goberman2022pdspeech}, these features were not consistently perceptible and varied substantially across speakers, even within the same severity level. In particular, several speakers with an MDS-UPDRS-S score of 2 exhibited heterogeneous speech characteristics, suggesting variability near the Mild–Severe boundary. Also, we found  that consistently misclassified speakers with Mild PD (score 0–1) showed little or no audible abnormalities, indicating that distinguishing HC from Mild PD remains particularly challenging for the classification pipelines. These finding are in line with the current challenges of diagnosing PD \cite{tolosa2021challenges}, in which individuals with other neurological conditions may be misdiagnosed as PD, as some PD labels may be incorrect, and some control participants may be in the PD prodromal phase.

Furthermore, comparing the original recordings with preprocessed segments revealed additional insights. Several speakers with an MDS-UPDRS-S speech score of 2 exhibited frequent unnatural silences or short, effortful speech bursts, and substantial variability in speech volume across datasets. However, segmentation via VAD and subsequent loudness normalization removed much of the silence structure and inter-speaker volume differences from the input data. Given that unnatural silences and reduced speech intensity are characteristic of PD speech \cite{goberman2022pdspeech}, the loss of these cues during preprocessing may have contributed to persistent misclassifications.

%

A key limitation of this work is that, while the same speech task was employed across three languages, each language was drawn from a different dataset; as a result, cross-lingual adaptation may have captured dataset-specific properties, such as recording conditions and channel variability, rather than purely linguistic differences \cite{botelho22_interspeech}. Second, multi-class severity modeling was based on binning MDS-UPDRS-S speech scores due to the limited availability of Severe PD speakers; while methodologically necessary, this discretization is less clinically grounded than the full continuous UPDRS scale and may obscure subtle gradations of disease severity. Finally, preprocessing decisions, including VAD-based segmentation and segment-level loudness normalization, likely attenuated prosodic and speech volume cues relevant to PD severity, potentially constraining the models' ability to capture clinically meaningful variation.

Future work should therefore evaluate preprocessing strategies that preserve prosodic information and inter-speaker intensity differences in the embeddings, while systematic variation of the k-shot target-language set size would clarify the necessary quantity of labelled target data for cross-lingual PD severity assessment. Furthermore, complementary explainable approaches could be investigated to potentially relate SFM embeddings to clinically interpretable acoustic features \cite{Gomez2025interpretability} \cite{zhong25recapd}.

\section{Conclusions}\label{sec:conclusion}

In this study, we have conducted the first systematic multi-class cross-lingual PD-severity assessment using open-source SFM embeddings. Our results showed that a k-shot adaptation strategy, in which the model is exposed to a small, balanced set of target-language segments, consistently improves accuracy and F1-score over zero-shot classification and can achieve strong performance on the target language without substantially degrading source-language results. However, outcomes remain strongly dependent on the dataset and severity class, particularly for the PC-GITA dataset and Severe PD class, indicating that data characteristics substantially influence performance. These results support cross-lingual training with SFM embeddings as an effective solution for limited labeled PD severity speech data, enabling robust cross-language transfer and scalable deployment in multilingual clinical settings.

Our work also revealed a subset of consistently hard-to-classify speakers, indicating that misclassifications are influenced by inter-speaker-specific patterns, potential label ambiguity near severity boundaries, and preprocessing steps that may attenuate clinically relevant cues such as pausing behavior and speech intensity.
Future work should prioritize the development of balanced, standardized multilingual PD-severity datasets and systematic cross-lingual and explainable modeling strategies to better align speech embeddings with clinically meaningful PD characteristics. 
Advancing robust and generalizable speech-based PD assessment will require the research community to rigorously evaluate cross-lingual SFM embeddings under clinically grounded severity modeling settings.

\section{AI-Generated Content Disclosure}
Generative AI tools were used for language editing and polishing, including grammar and phrasing. All scientific content, experimental design, analyses, results, and conclusions were developed, verified, and approved by the authors. The authors take full responsibility for the content of this paper, and no generative AI tool is listed as a co-author.

\bibliographystyle{IEEEtran}
\bibliography{mybib}

\end{document}